\documentclass[prl,twocolumn,fleqn]{revtex4}
\usepackage{epsfig}
\usepackage{epstopdf}
\usepackage{graphicx}

\usepackage{amsmath}
\usepackage[psamsfonts]{amssymb}
\usepackage{euscript}

\usepackage{latexsym}

\def\hsp{,\hspace{.7cm}}

\def\thet#1#2{\theta_{#1#2}}

\def\sp#1#2#3{\hbox{\rm sin}^#1(\thet#2#3)}

\def\spp#1#2#3#4{\hbox{\rm sin}^#1(#2\thet#3#4)}

\def\m#1#2{\ensuremath{\Delta M_{#1#2}^2}}

\renewcommand{\sin}{\textrm{sin}}

\renewcommand{\(}{\begin{equation}}
\renewcommand{\)}{end{equation} \vspace{-.05in}\linebreak}

\newcounter{saveeqn}
\newcounter{savealpheqn}

\newcommand{\alpheqn}{\setcounter{saveeqn}{\value{equation}}%
  \stepcounter{saveeqn}\setcounter{equation}{0}%
  \renewcommand{\theequation}{\mbox{\arabic{section}.\arabic{saveeqn}
\alph{equation}}}
  \renewcommand{\)}{\end{equation}}}
\def\part#1{\frac{\partial}{\partial{#1}}}%
\def\group#1{\refstepcounter{equation}\setcounter{saveeqn}
 {\value{equation}}%
  \label{#1}\setcounter{equation}{0}%
\renewcommand{\theequation}{\mbox{\arabic{section}.\arabic{saveeqn}
\alph{equation}}}
  \renewcommand{\)}{\end{equation}}}
\newcommand{\reseteqn}{\setcounter{equation}{\value{saveeqn}}%
  \renewcommand{\theequation}{\arabic{section}.\arabic{equation}}%
  \renewcommand{\)}{\end{equation}}}

\newcommand{\aalpheqn}{\setcounter{saveeqn}{\value{equation}}%
  \stepcounter{saveeqn}\setcounter{equation}{0}%
  \renewcommand{\theequation}{\mbox{
        \Alph{subsection}.\arabic{saveeqn}\alph{equation}}}
   \renewcommand{\)}{\end{equation}}}
\newcommand{\areseteqn}{\setcounter{equation}{\value{saveeqn}}%
  \renewcommand{\theequation}{\Alph{subsection}.\arabic{equation}}%
  \renewcommand{\)}{\end{equation}}}

\renewcommand{\thefootnote}{\alph{footnote}}
\renewcommand{\(}{\begin{equation}}
\renewcommand{\)}{\end{equation}}
\newcommand{\ba}{\begin{eqnarray}}
\newcommand{\ea}{\end{eqnarray}}

\newcommand{\bp}{\mathop{\vtop{\ialign{##\crcr
   $\hfil\displaystyle{}\hfil$\crcr\noalign{\kern-13pt\nointerlineskip}
   \BIG{(}\hskip0pt\crcr\noalign{\kern3pt}}}}}
\newcommand{\cbp}{\mathop{\vtop{\ialign{##\crcr
   $\hfil\displaystyle{}\hfil$\crcr\noalign{\kern-13pt\nointerlineskip}
   \BIG{)}\hskip0pt\crcr\noalign{\kern3pt}}}}}
\newcommand{\pa}{\mathop{\vtop{\ialign{##\crcr
    
$\hfil\displaystyle{\oplus}\hfil$\crcr\noalign{\kern+1pt\nointerlineskip 
}
   \hspace{.08in}$^{\alpha=0}$\hskip6pt\crcr\noalign{\kern3pt}}}}}
\renewcommand{\hsp}{,\hspace{.3in}}

\newcommand{\beq}{\begin{equation}}
\newcommand{\eeq}{\end{equation}}

\numberwithin{equation}{section}
\renewcommand{\theequation}{\mbox{\arabic{equation}}}

\def\hsp#1{\hspace{#1in}}

\catcode`\@=11
\def\vereq#1#2{\lower3pt\vbox{\baselineskip1.5pt \lineskip1.5pt
\ialign{$\m@th#1\hfill##\hfil$\crcr#2\crcr\sim\crcr}}}
\catcode`\@=12

\makeatletter
\newcommand\figcaption{\def\@captype{figure}\caption}
\newcommand\tabcaption{\def\@captype{table}\caption}
\makeatother
\renewcommand{\(}{\begin{equation}}
\renewcommand{\)}{\end{equation}}

\def\m#1#2{\ensuremath{\Delta M_{#1#2}^2}}

\renewcommand{\beq}{\begin{equation}}
\renewcommand{\eeq}{\end{equation}}
\newcommand{\bea}{\begin{eqnarray}}
\newcommand{\eea}{\end{eqnarray}}
\newcommand{\beas}{\begin{eqnarray*}}
\newcommand{\eeas}{\end{eqnarray*}}

\newcommand{\bquo}{\begin{quote}}
\newcommand{\enqu}{\end{quote}}

\def\hsp{,\hspace{.2cm}}

\begin{document}
\def\thefootnote{\fnsymbol{footnote}}

\title{The Leptonic CP Phase from Muon Decay at Rest with Two Detectors}

\author{Emilio Ciuffoli, Jarah Evslin and Xinmin Zhang}
\affiliation{Theoretical Physics Division, IHEP, 
CAS, YuQuanLu 19B, Beijing 100049, China 
}

\begin{abstract}
\noindent

\noindent
We propose a novel experimental setup for the determination of the  leptonic CP-violating phase $\delta$ using the decay at rest (DAR) of $\mu^+$ from a single source located at distances of 10 and 30 km from  two 20 kton organic liquid scintillator detectors.  The $\mu^+$ are created by bombarding a target with a 9 mA beam of 800 MeV protons. With this proposal $\delta$ can be determined with a precision of about 20 (15) degrees in 6 (12) years.  In contrast with the DAE$\delta$ALUS project, only a single source is required and it runs with a duty factor of 100\%.  Therefore 9 mA is the maximum instanteous current, greatly reducing both the technological challenges and the costs.




\end{abstract}

%
\setcounter{footnote}{0}
\renewcommand{\thefootnote}{\arabic{footnote}}


\maketitle

The CP violation in the standard model is not sufficient to explain the observed matter-antimatter asymmetry.   In many models this asymmetry is caused by CP violation in the leptonic sector, whose simplest origin is the unique phase $\delta$ which arises for Dirac neutrinos.  So far the value of $\delta$ is entirely unknown, however in the next decade the experiments NO$\nu$A and T2K will have some sensitivity to sin$(\delta)$, although they will not be able to distinguish $\delta$ from $\pi-\delta$ \cite{noiinterf,parkedegen}.  Future proposals in general are expensive and depend on unproven technology, such as the scalability of liquid argon detectors and the control of excitations of $H_2^+$ ions.  Our proposal will yield a 20 (15) degree precision measurement of $\delta$ in 6 (12) years using technology not far beyond the current state of the art and at a much lower cost than its competitors.




A cyclotron complex, consisting of a pair of cyclotrons, will accelerate protons to 800 MeV which then strike a target, producing $\pi^+$ that decay at rest.  The resulting $\mu^+$ will in turn decay at rest creating $\overline{\nu}_\mu$ that then oscillate to $\overline{\nu}_e$.  These $\overline{\nu}_e$ are detected via inverse $\beta$ decay by two organic liquid scintillator detectors, each with a target mass of 20 kton and consisting of 12\% free protons, located 10 and 30 km from the complex.   Our proposal requires a maximum instantaneous proton current which is appreciably lower than that required by the DAE$\delta$ALUS experiment, greatly reducing the technological requirements on the cyclotrons.  



As has been proposed in Ref.~\cite{2rivel}, such a pair of detectors may be employed by the JUNO \cite{juno} and RENO 50 \cite{reno50} experiments to determine the neutrino mass hierarchy using reactor neutrinos.  The employment of a pair of detectors eliminates the loss of sensitivity to mass hierarchy which would otherwise result from the detector's unknown nonlinear energy response.  The mass hierarchy and CP violation experiments may be performed simultaneously.

The manifestation of the DAE$\delta$ALUS proposal in Ref.~\cite{daed} and the variations which have since been proposed \cite{lena,whitepaper} each require three cyclotron pairs, each of which costs at least 25 to 100 million dollars \cite{whitepaper}.  Thus our proposal, with a single complex, represents a significant savings.  The liquid scintillator detectors cannot determine the angle from which the neutrinos arrived and the beams are not pulsed, therefore only one cyclotron may run at a time.   This is particularly problematic as these proposals are always statistics limited, even when the detector is as large as hyperK \cite{whitepaper}.  Thus our proposal, with a single $\mu^+$ source, enjoys the same neutrino flux as would a proposal with three sources.  In fact the signal is doubled as both detectors run simultaneously.  Moreover, the precision can be improved by running the experiment for longer, by increasing the target mass or, if more funds become available, by adding more cyclotron complexes close to the original complex, so that they may run simultaneously.

The determination of $\delta$ using $\mu^+$ decay at rest has several advantages.  For one, the expected neutrino spectrum is known precisely.  Second, in the window between 20 MeV and 55 MeV the $\overline{\nu}_e$ backgrounds are extremely low and, as the detectors in our proposal are liquid scintillators, there is no invisible muon background.  Third, as this energy range is distinct from the 2-8 MeV energy range of the reactor neutrinos used by these experiments to determine the mass hierarchy, the reactor and $\mu^+$ decay at rest experiments may run simultaneously.  Finally, this determination uses antineutrino oscillations, which have maximum synergy with accelerator experiments like T2K, NO$\nu$A and LBNE that enjoy better statistics in the neutrino oscillation channel.  More precisely, by  comparing the two channels one can remove the degeneracy between $\delta$, which can be extracted from the difference between the appearance in the neutrino and antineutrino channels, and a combination of $\theta_{23}$ and $\theta_{13}$, which increases the appearance in both channels simultaneously and so can be extracted from the sum of the appearance rates.   We will leave a computation of the combined sensitivity of this proposal with an accelerator experiment to future work.

\begin{figure} 
\begin{center}
\includegraphics[width=2.8in,height=1.2in]{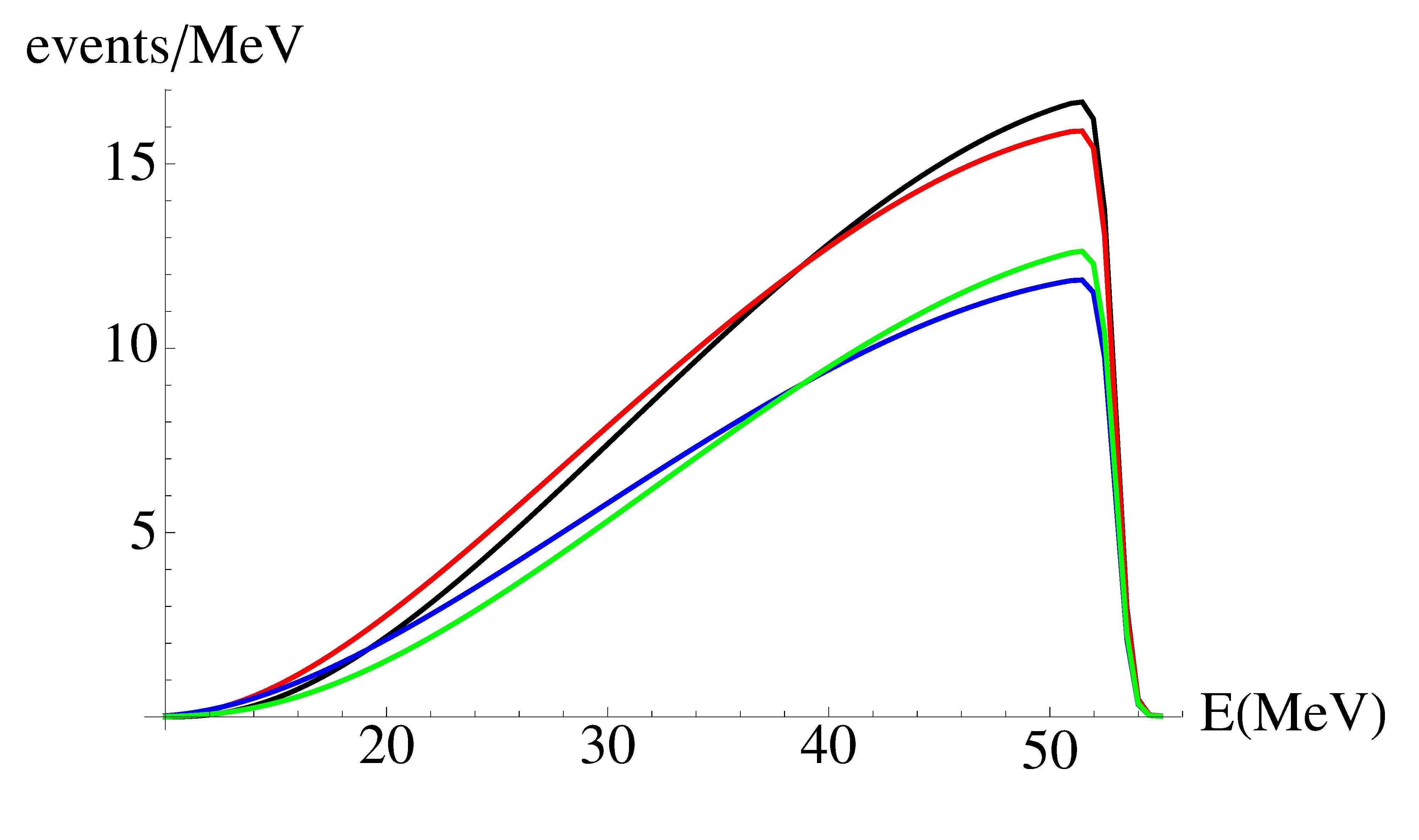}
\includegraphics[width=2.8in,height=1.2in]{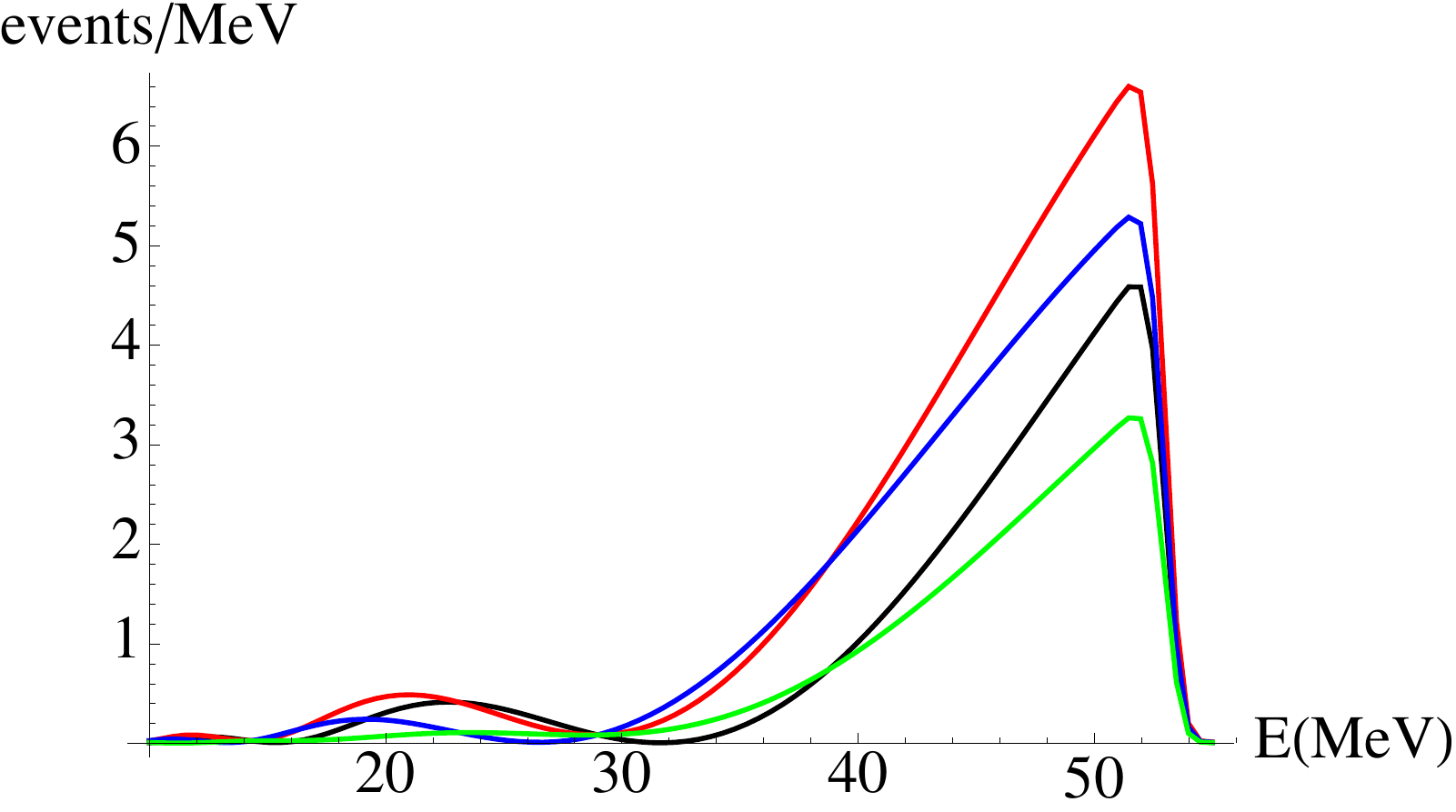}
\caption{Number of $\overline{\nu}_e$ per MeV expected at the near detector (top) and the far detector (bottom) in 6 years if $0^\circ$ (black), $90^\circ$ (red), $180^\circ$ (blue) and $270^\circ$ (green).}
\label{spettra}
\end{center}
\end{figure}

In the calculations below we have fixed the normalization of the number of IBD events such that,
 with $\delta=0$, at 10 km a 20 kton target mass detector, consisting of $12\%$ free protons, will observe $350$
 events in $6$ years.  Scaling the $\mu^+$ decay rate at LSND \cite{lsnd}, this corresponds to about $45$ MW
 years of power using a $800$ MeV proton beam, or equivalently $1.8\times 10^6$ C of protons.  For example,
 for a $6$ year run one would require a constant current of $9$ mA.    The critical advantage of our proposal is
 that the proton beam runs with a duty factor of essentially $100\%$, so $9$ mA is not only the average current, but also
 the peak current.  This is a factor of $4$ less than the peak current required in phase two of DAE$\delta$ALUS
 \cite{daed,daedseminario}, greatly reducing the technological requirements on the cyclotron.  For example, the cyclotron
 proposed in Ref.~\cite{atomic}
 would be sufficient.  
For simplicity below we will assume a constant current of $9$ mA and report the number of years of running.  However our results can easily be generalized to different currents and target masses by simply linearly scaling the livetime. 

As explained in Ref.~\cite{whitepaper} a factor of 2 in the beam power at fixed perveance may be gained by accelerating $H_2^+$ ions.  This is challenging as the $H_2^+$ excited states need to be controlled.  As a result of our lower beam power requirements, a proton beam may well be sufficient for this proposal.

In our simulations we have restricted our attention to the normal neutrino mass hierarchy, corresponding to the assumption that the hierarchy will be known before this experiment takes place.   Similarly we considered only the tree level IBD cross section and ignored neutron recoil.  While these effects do need to be considered in the fitting of $\delta$ in the true experiment, their inclusion in the simulation and fitting procedure would not significantly affect the precision with which $\delta$ can be determined.  However we have included matter effects as the neutrinos travel through the Earth. 

We fix the neutrino mass differences to be
\beq
\m31=2.4\times 10^{-3}{\rm{eV}}^2\hsp
\m21=7.5\times 10^{-5}{\rm{eV}}^2
\eeq
and the neutrino mass mixing angles to be
\beq
\spp2213=0.089,\ 
\spp2212=0.857,\ 
\sp223=\frac{1}{2}.
\eeq
The expected spectra at the near (10 km) and far (30 km) detectors after a 6 year run are plotted in Fig.~\ref{spettra} for $\delta=0^\circ,\ 90^\circ,\ 180^\circ$ and $270^\circ$.  Note that the differences in shape of the spectra are fairly small, but that the energy resolution of a liquid scintillator detector is quite good in this energy range.  In fact JUNO and RENO 50, in part because of the density and quality of their PMTs, are expected to have an energy resolution as much as a factor of two better than LENA.   We used a somewhat conservative fractional energy resolution
\beq
\frac{\delta E}{E}=\sqrt{\left(\frac{3\%}{\sqrt{E/{\rm{MeV}}}}\right)^2+\left(1\%\right)^2} .
\eeq

\begin{figure} 
\begin{center}
\includegraphics[width=2.8in,height=1.2in]{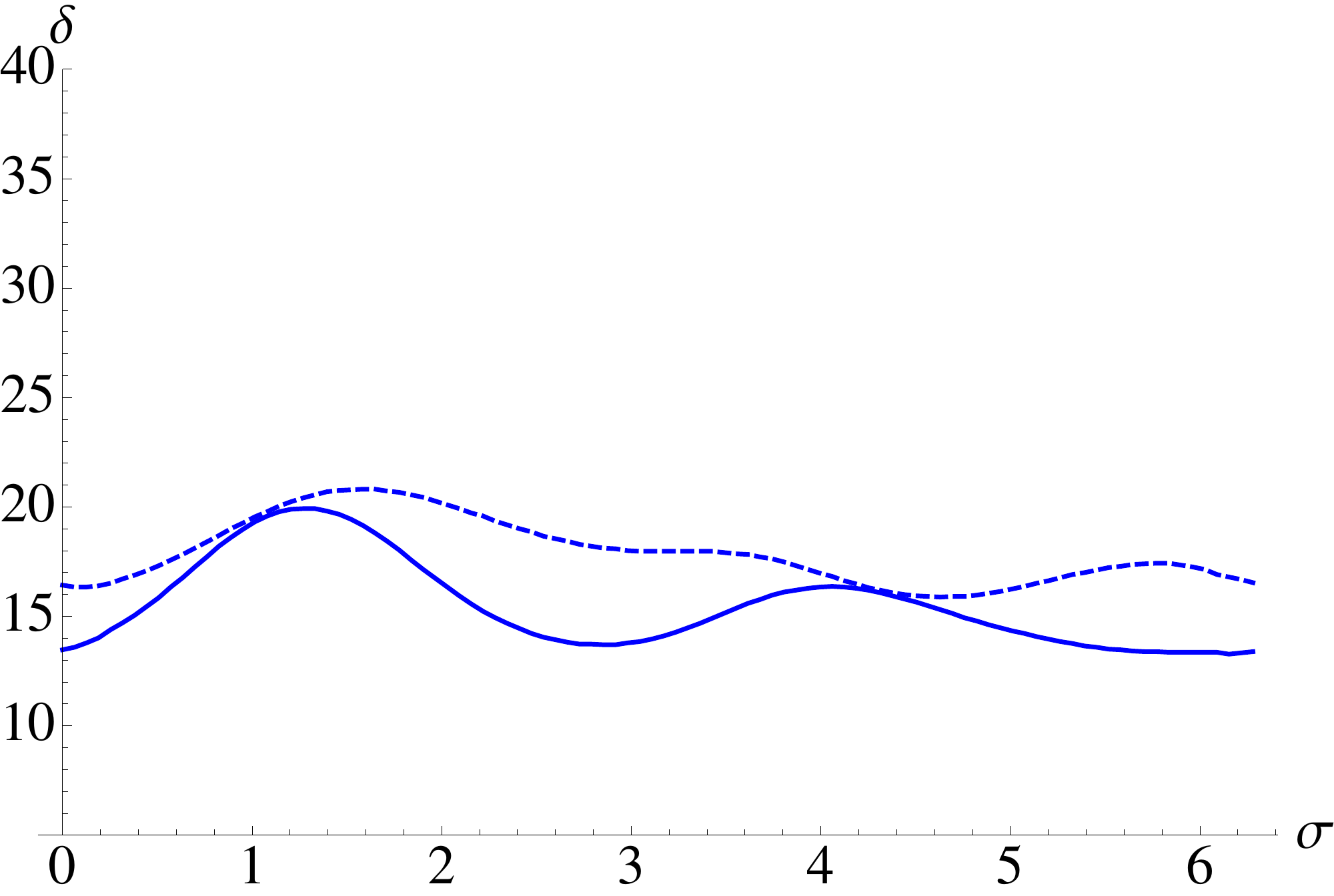}
\includegraphics[width=2.8in,height=1.2in]{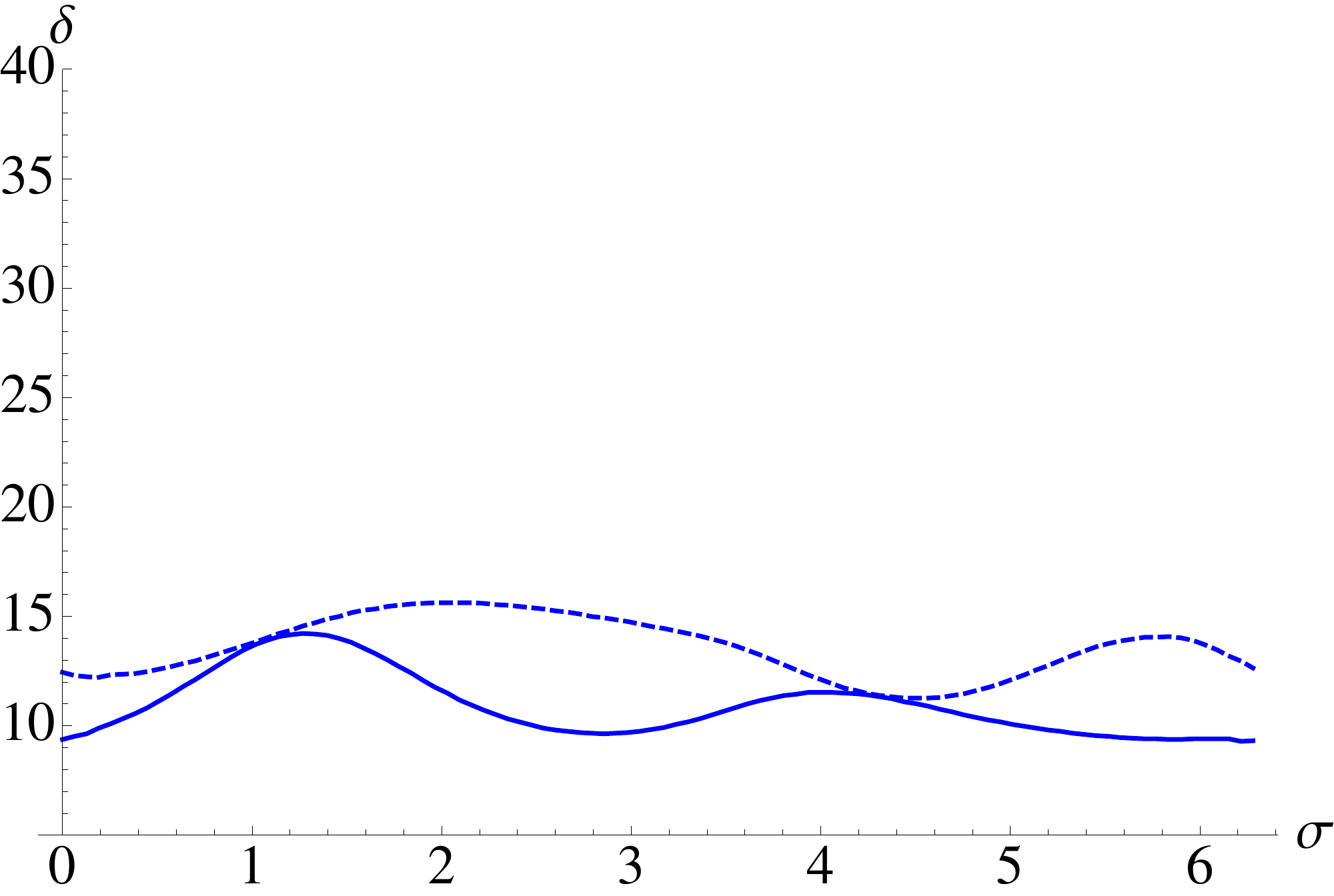}
\caption{The 1$\sigma$ precision, in degrees, with which $\delta$ can be determined for various values of $\delta$ in a 6 year (top) and 12 year (bottom) run.  In this figure we have assumed that the mass matrix mixing angles are known perfectly.  The total normalization of the neutrino flux is known perfectly (solid curve) and with a precision of 5\% (dashed).}
\label{perfetto}
\end{center}
\end{figure}

We determine the $1\sigma$ precision with which these experiments may determine $\delta$ using a Poisson-statistics $\chi^2$ fit to the Asimov data set.  In Fig.~\ref{perfetto} we report this precision for 6 and 12 years of running.  In this figure we assume that the mixing angles are known perfectly and we consider both the case in which the flux normalization is known perfectly and also an uncertainty in the flux normalization of 5\%, which is treated using the standard pull parameter method of Ref.~\cite{pull}.   To test these results we have also performed a series of Monte Carlo simulations for each mixing angle.  The results of our Monte Carlo are compatible with those of the $\chi^2$ analysis presented in Fig.~\ref{perfetto}.

\begin{figure} 
\begin{center}
\includegraphics[width=2.8in,height=1.2in]{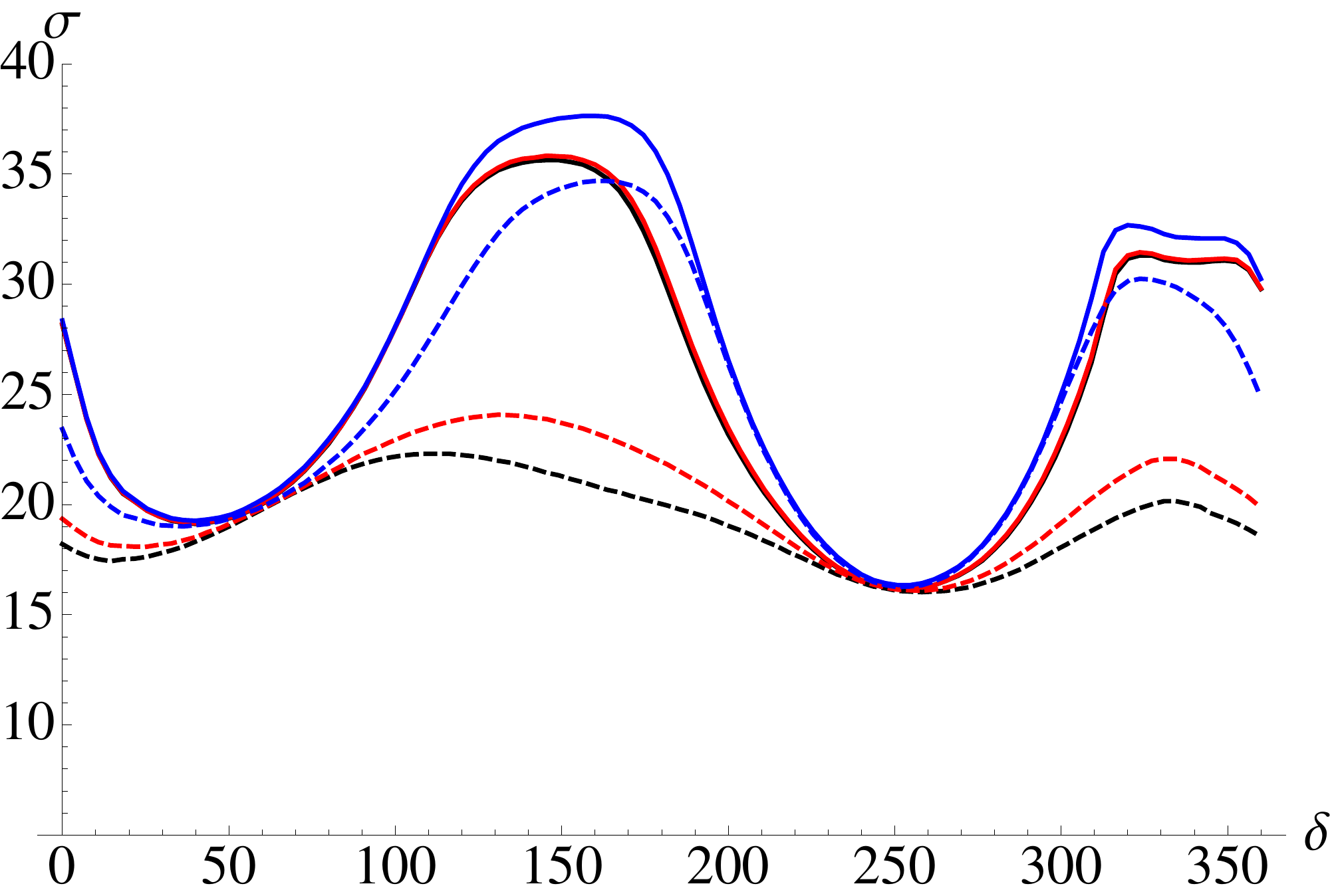}
\includegraphics[width=2.8in,height=1.2in]{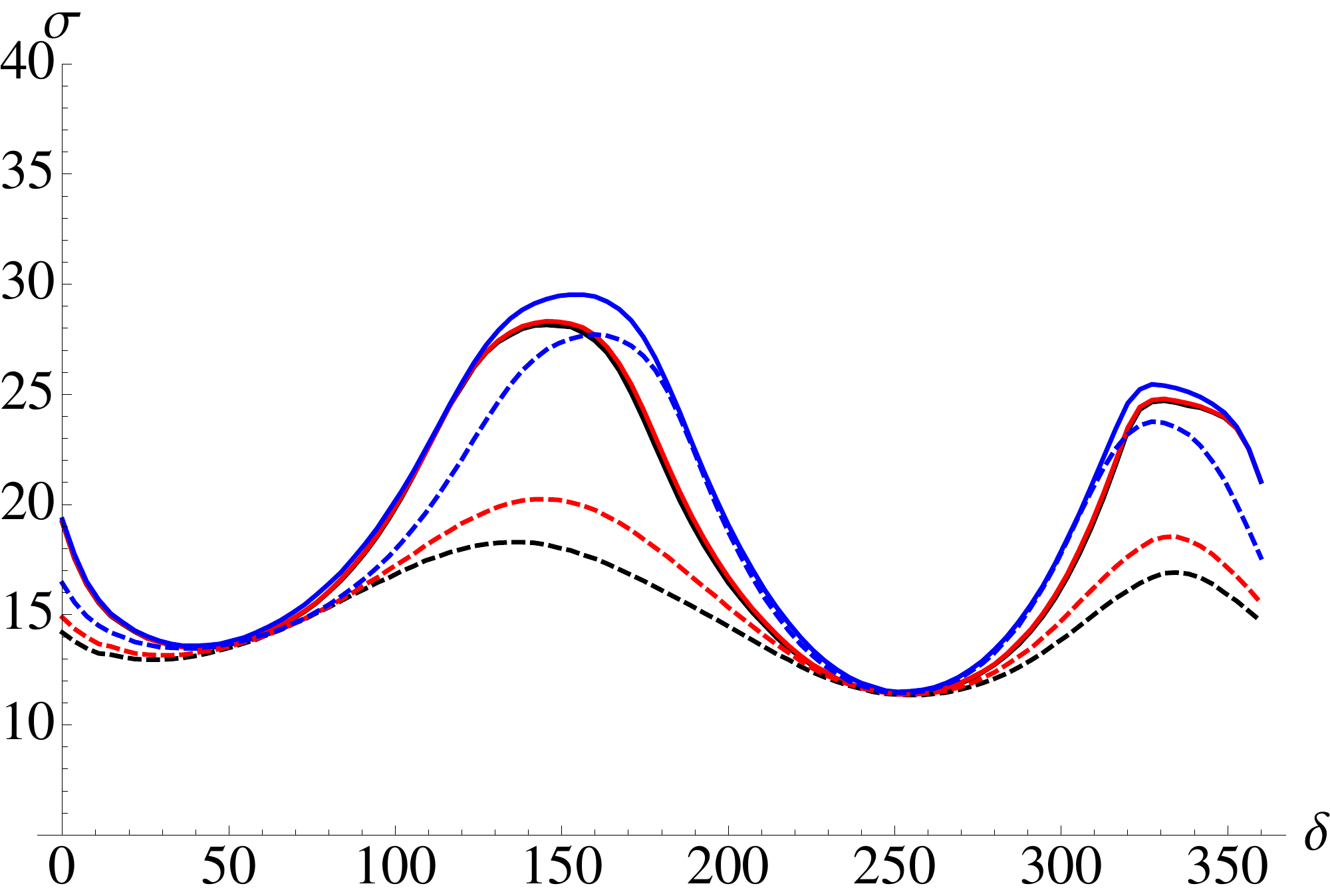}
\caption{The 1$\sigma$ precision, in degrees, with which $\delta$ can be determined for various values of $\delta$ in a 6 year (top) and 12 year (bottom) run.   The total normalization of the neutrino flux is known with a precision of 1\% (black), 5\% (red) and 20\% (blue). The solid and dashed curves correspond to current and future uncertainties in the mixing angles.}
\label{nuovavec}
\end{center}
\end{figure}

\begin{figure} 
\begin{center}
\includegraphics[width=2.8in,height=1.2in]{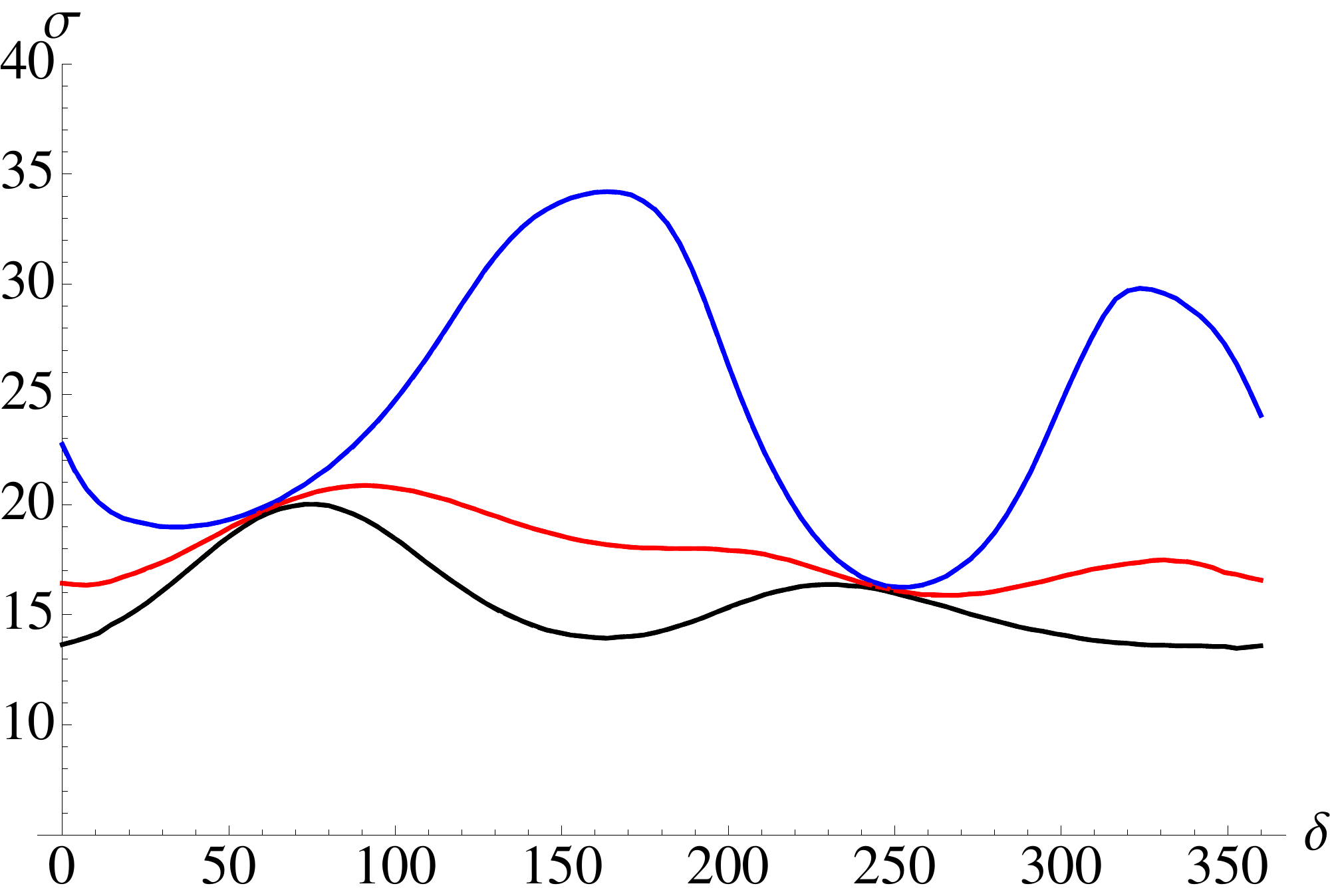}
\includegraphics[width=2.8in,height=1.2in]{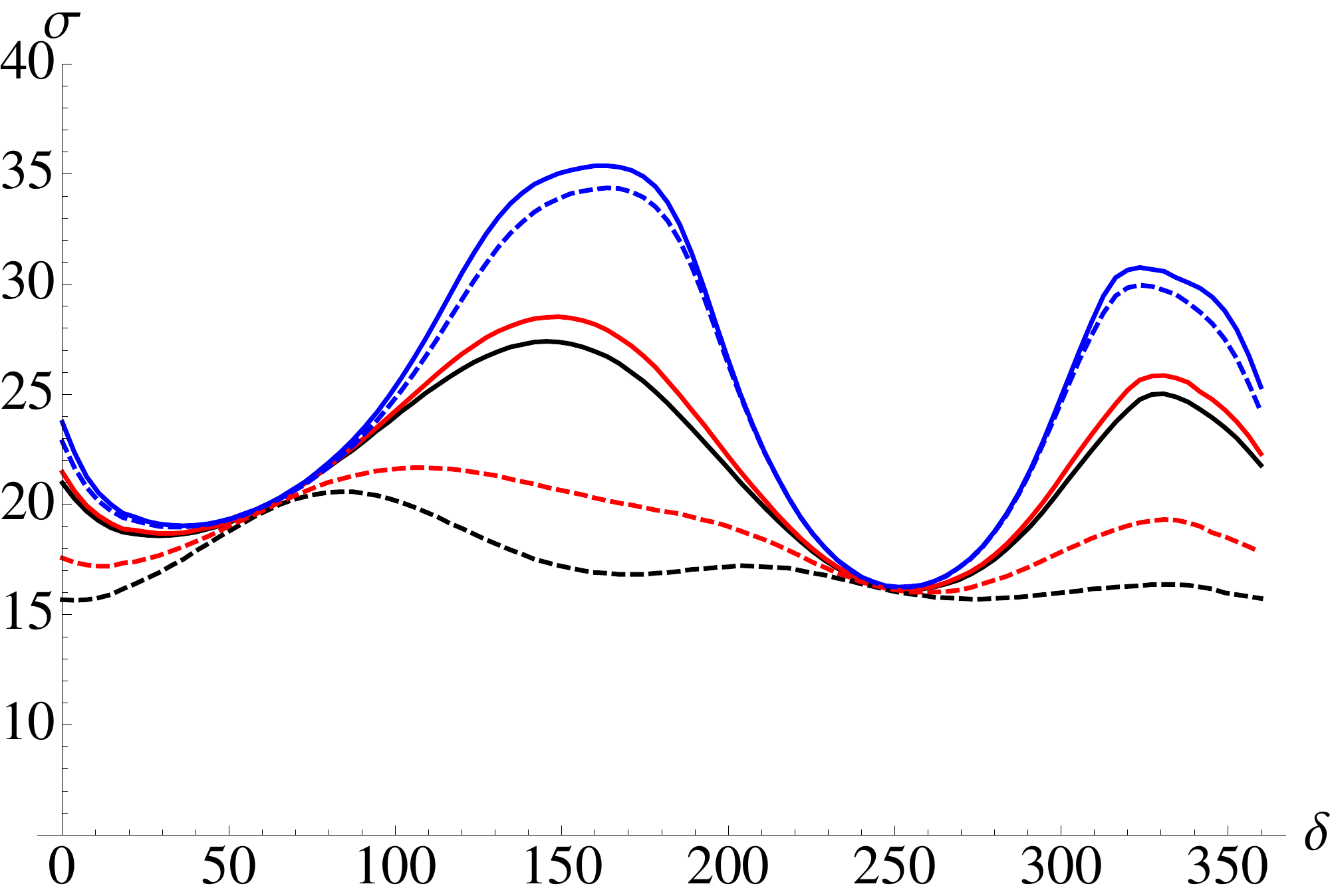}
\includegraphics[width=2.8in,height=1.2in]{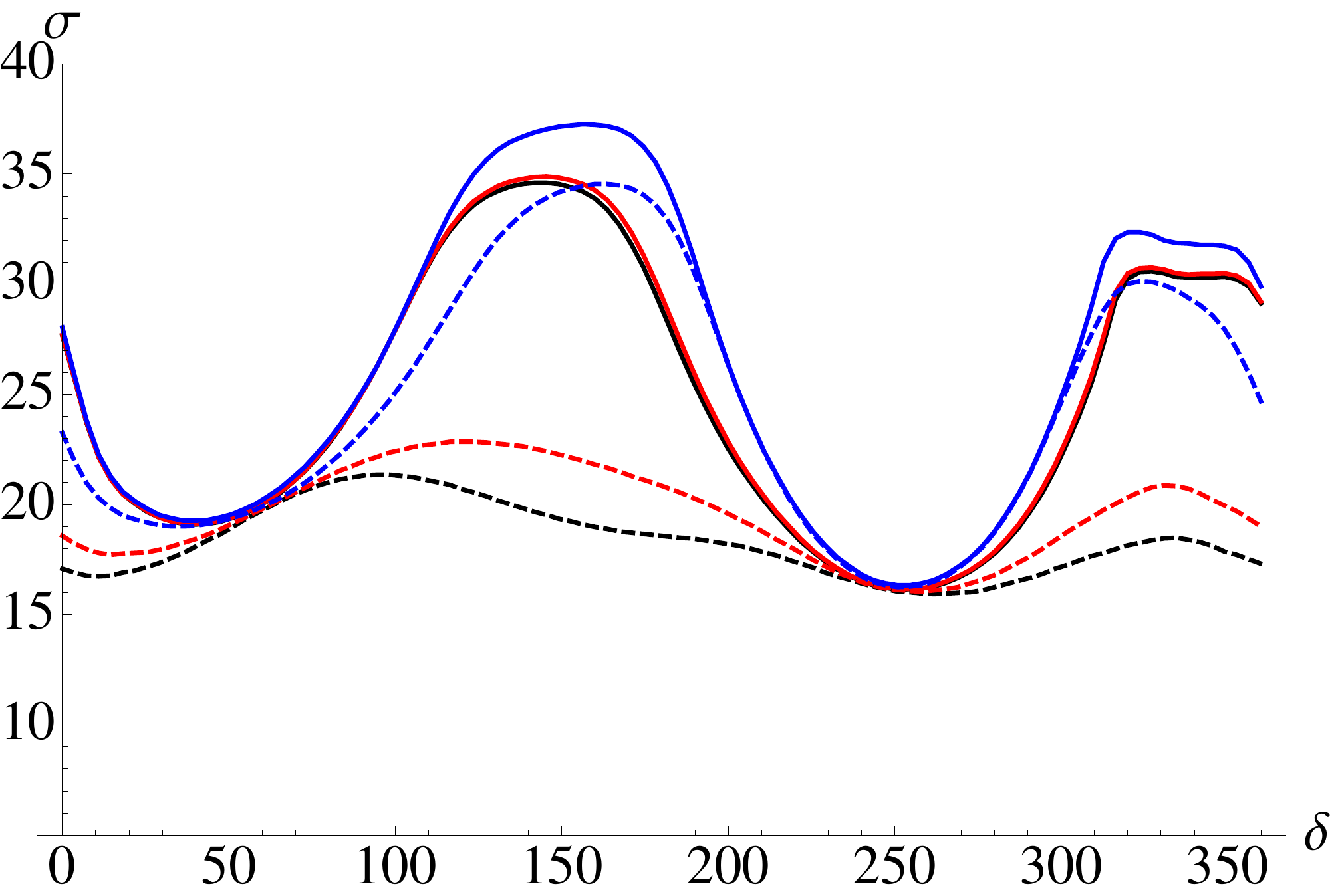}
\caption{Six years of running, as in Fig.~\ref{nuovavec} except that  all of the mixing angles are fixed except for $\theta_{12}$ (top), $\theta_{13}$ (middle) and $\theta_{23}$ (bottom).  Note that the error in $\theta_{12}$ has essentially no effect on the precision with which $\delta$ can be determined.} 
\label{unangolo}
\end{center}
\end{figure}

Of course the mixing angles will not be known perfectly.   In Fig.~\ref{nuovavec} we include uncertainties in the mixing angles corresponding to the current uncertainties
\bea
&&\delta\spp2212=0.024\hsp \delta\spp2213=0.01\nonumber\\
&& \frac{\delta\sin(\theta_{23})}{\sin(\theta_{23})}=11\%
\eea
and also with uncertainties expected when experiments currently running are finished
\bea
&&\frac{\delta\spp2212}{\spp2212}=1\% \hsp \frac{\delta\spp2213}{\spp2213}=4\%\nonumber\\
&& \delta\sin(\theta_{23})=0.02.
\eea
To understand the relevant contributions of the uncertainties from the different mixing angles, in Fig.~\ref{unangolo} we have fixed all of the angles except for one, to which we have applied the current and future uncertainties.  The main contribution to the uncertainty comes from a single combination of $\theta_{13}$ and $\theta_{23}$, this degeneracy will be broken by combining data from $\mu^+$ decay at rest with the neutrino appearance channel from accelerator experiments.

Our main result is Fig.~\ref{nuovavec}.  As can be seen, a single cyclotron complex, producing antineutrinos using $\mu^+$ decay at rest, can determine $\delta$ with a precision of 20 (15) degrees in 6 (12) years using detectors that may anyway be built for reactor neutrino experiments.  Backgrounds are expected to be small in this energy range and studies of such experiments \cite{whitepaper} have consistently shown, albeit with 3 cyclotron complexes, that systematic errors are extremely small.  Nonetheless in a subsequent publication we will examine the effects of these systematic errors and backgrounds as well as an optimization of the location of the cyclotron complex.

\begin{figure} 
\begin{center}
\includegraphics[width=2.8in,height=1.2in]{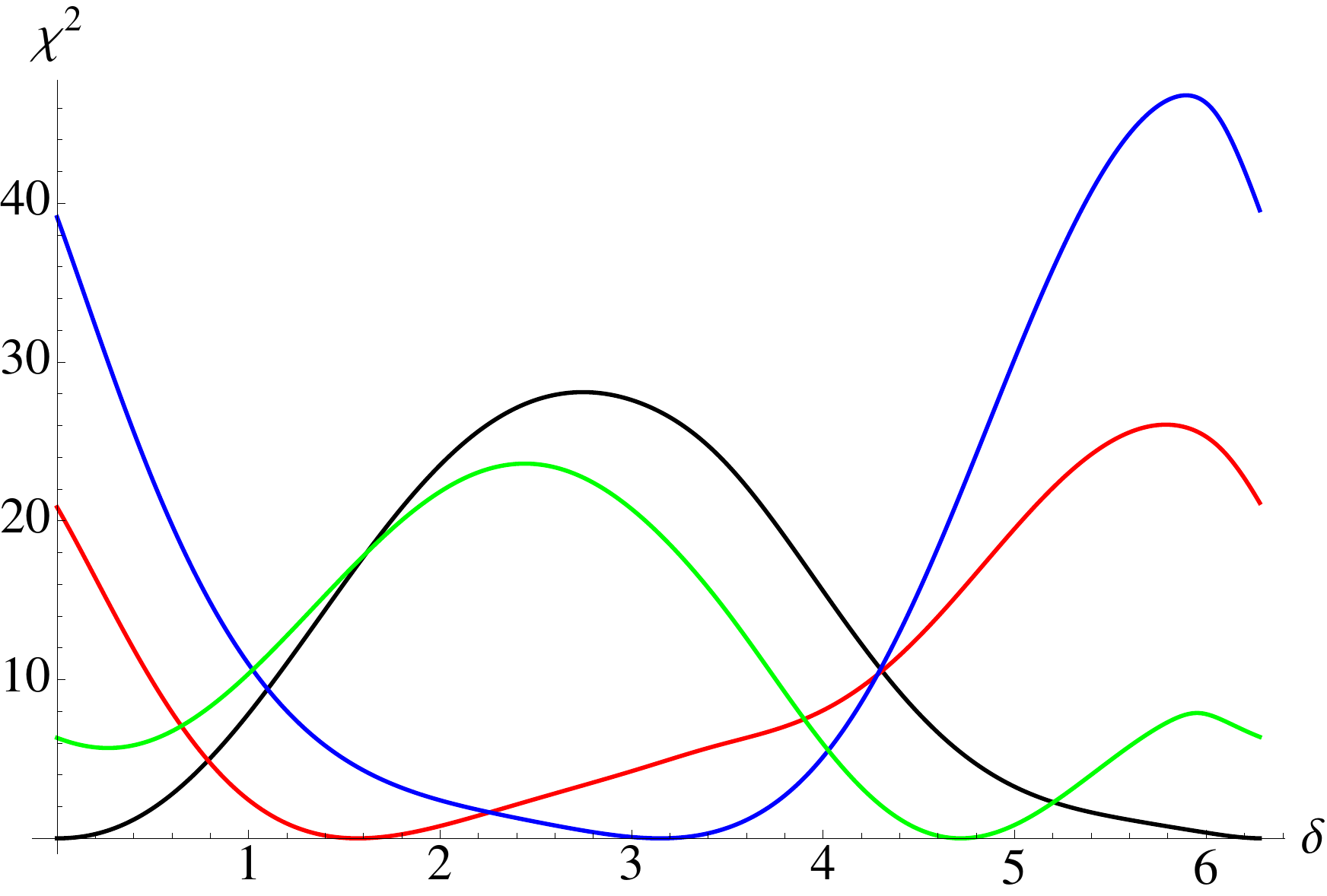}
\includegraphics[width=2.8in,height=1.2in]{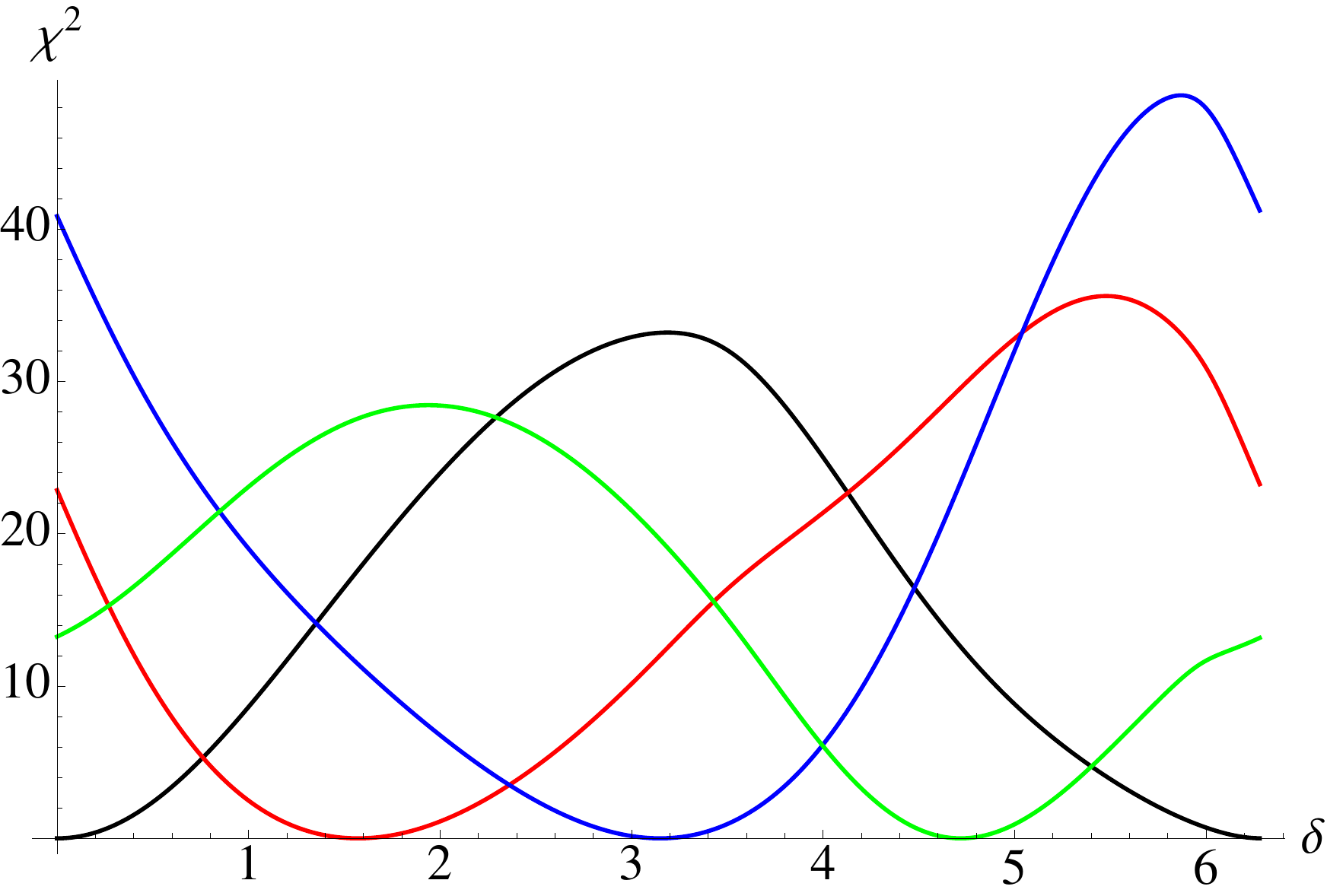}
\caption{The true values of $\delta$ are $0^\circ$ (black), $90^\circ$ (red), $180^\circ$ (blue) and $270^\circ$ (green).  The reported values of $\chi^2$ correspond to fits of 6 years of data to various values of $\delta$ assuming a 5\% uncertainty in the normalization and the current (top panel) and future (bottom panel) uncertainties in the mixing angles.  Note that there is essentially no degeneracy between $\delta$ and $180^\circ-\delta$.} 
\label{nodegen6}
\end{center}
\end{figure}

In Refs.~\cite{noiinterf,parkedegen} it was noted that off axis accelerator experiments designed to measure $\delta$ have relatively monochromatic beams and so are sensitive primarily to the flux at the oscillation maximum, which depends only upon $\sin(\delta)$ and so cannot distinguish $\delta$ from $180^\circ-\delta$.  On the other hand, the muon decay at rest spectrum is far from monochromatic and so there is no such degeneracy in the $\delta$ determined by such experiments.  This can be seen in Fig.~\ref{nodegen6} where $\chi^2$ is plotted as a function of the $\delta$ in the fitting function, the degeneracy would correspond to a local minimum at the $180^\circ$ minus the true value of $\delta$.  No such minimum is present in the figures.

\section* {Acknowledgement}

\noindent
We have benefited  from discussions with Jianjun Yang and Tianjue Zhang.   It is our pleasure to thank Shao-Feng Ge for finding an error in an earlier version of this manuscript.  JE is supported by NSFC grant 11375201.  EC  is supported by the Chinese Academy of Sciences Fellowship for Young International Scientists grant  number 2013Y1JB0001 and NSFC grant 11350110500.  XZ is supported in part by  NSFC grants 11121092, 11033005 and 11375202.   


\end{document}